\documentclass[
aps, pra, reprint
]{revtex4-2}

\usepackage[english,ngerman]{babel} 
\usepackage[utf8]{inputenc}

\usepackage{amsmath,amssymb,amsthm} 
\usepackage{mathtools}              
\usepackage{microtype}              
\usepackage{bm}                     
\usepackage{dcolumn}                

\usepackage{graphicx}
\usepackage[version=4]{mhchem}
\usepackage{qcircuit}
\usepackage{tabularx}
\usepackage{colortbl}
\usepackage{diagbox}
\usepackage{placeins}

\usepackage{algorithm}
\usepackage{algpseudocode}
\floatname{algorithm}{Protocol}

\usepackage{xcolor}
\usepackage{listings}
\usepackage{braket}
\usepackage[most]{tcolorbox}

\newtheorem{theorem}{Theorem}

\theoremstyle{definition}

\theoremstyle{remark}

\usepackage{algorithm}
\floatname{algorithm}{Algorithm}
\usepackage{algpseudocode}

\definecolor{myrefcolor}{rgb}{0.067,0.5,0.5}
\usepackage[
  pdftex,
  breaklinks,
  colorlinks=true,
  linkcolor=myrefcolor,
  citecolor=myrefcolor,
  urlcolor=myrefcolor
]{hyperref}

\begin{document}

\title{Krylov Polynomials and Quantum Query Complexity}

\author{Kiran Adhikari}
\email{kiran.adhikari@tum.de}
\affiliation{Emmy Noether Group for Theoretical Quantum Systems Design, Technical University of Munich, Germany}


\begin{abstract}
We show that the minimal query complexity for preparing 
$f(H)\ket{\psi_0}$ is exactly the optimal polynomial approximation 
degree of $f$ in $L^2(\mu)$, where $\mu$ is the spectral measure of 
$(H,\ket{\psi_0})$. This state-aware perspective refines the worst-case 
bounds, unifies Krylov/Favard approximation with quantum queries, and 
explains how state-dependent spectral structure can yield 
substantial savings over uniform designs.

\end{abstract}

\maketitle


\section{Introduction}
In many quantum algorithms, prior information about the initial state is available or can be structurally assumed. When an algorithm explicitly leverages such information, particularly the spectral measure of the state with respect to the Hamiltonian, we describe the approach as utilizing a state-aware formalism. Prominent examples include quantum linear system algorithms (QLSAs) such as HHL, eigenvalue estimation, quantum walks from a localized source, and spectral filtering problems \cite{Harrow_2009, Venegas_Andraca_2012, Kempe_2003}. In the state-aware formalism, the effective dynamics are captured not by the full Hilbert space, but by the considerably smaller Krylov subspace.

Recently, Krylov methods have attracted interest in the quantum domain in the context of quantum chaos and various complexity measures \cite{M_ck_2022, Nandy_2025, rabinovici2025krylovcomplexity,Balasubramanian_2022, Adhikari_2023,Adhikari_2024,balasubramanian2025variationsthemekrylov, caputa2025symmetryresolvedspreadcomplexity}. Similarly, Krylov techniques have also been extensively employed for quantum simulation problems \cite{kirby2023exact, Motta_2019, kirby2025quantumkrylovalgorithmszego, jamet2021krylovvariationalquantumalgorithm, PhysRevB.102.094315}.

One open problem is whether a relationship exists between Krylov-based complexity measures and quantum computational complexity. Certain correspondence has been developed between Nielsen's geometric complexity \cite{Nielsen_2006} and Krylov complexity by choosing the Krylov basis as a part of elementary gate sets in Nielsen geometry \cite{Craps_2024, craps2025explicitconnectionskrylovnielsen}. However, Krylov complexity itself is dependent on the initial state and time, thus cannot be directly related to circuit complexity a priori. In this paper, rather than the Krylov complexity, we show that an exact relationship exists between the Krylov space dimension and the quantum query complexity. For certain ground state preparation problems, circuit depth has been shown to be proportional to the Krylov dimension \cite{Motta_2019, kirby2023exact}. Using optimal orthogonal approximation techniques \cite{Chihara2011}, we generalize this observation for an arbitrary function $f(H)$. 

The central theme of our proposal is to replace the minimax (Chebyshev) polynomials in Quantum Singular Value Transformation (QSVT) framework \cite{Gily_n_2019, Martyn_2021} with the state-aware polynomials, which we call Krylov polynomials. In fact, polynomial methods have been widely used in obtaining lower bounds of quantum-query complexity \cite{beals1998quantumlowerboundspolynomials,ambainis2023exponentialseparationquantumquery}. The central intuition of our proposal is that knowing the input state is confined to a specific spectral support allows for significantly shallower circuits compared to worst-case uniform approximations. Then, the problem statement goes as follows. The goal is to prepare a state of the form
\begin{equation}
    |\psi_{\text{out}}\rangle = f(H)\,|\psi_0\rangle,
\end{equation}
where \( H \) is a Hermitian operator accessible via oracle or block-encoding queries, 
\( f \) is a target function, 
and \( |\psi_0\rangle \) is a given initial state. 
Some examples include time evolution $f(H) = e^{-iHt}$, imaginary time evolution $f(H) = e^{-\beta H}$, preparing a ground state or solving a linear problem $f(H) = H^{-1}$. 

Conventional analyses are state-oblivious and neglect the actual spectral support of the initial state, which leads to overly pessimistic bounds. However, repeated applications of \(H\) on the initial state generate the Krylov subspace
\[
\mathcal{K}_m(H,|\psi_0\rangle) = \mathrm{span}\{|\psi_0\rangle, H|\psi_0\rangle, \ldots, H^{m-1}|\psi_0\rangle\},
\]
within which all states of the form \( f(H)|\psi_0\rangle \) reside. 
A quantum circuit making \(q\) oracle queries to \(H\) can prepare only states 
\( p(H)|\psi_0\rangle \) where \(p\) is a polynomial of degree at most \(q\);  thus, we show that the query count and the Krylov depth are equivalent notions. More precisely, we show that the given full knowledge of the spectral measure $\mu_{H,\psi_0}$, the minimal query complexity equals the optimal polynomial approximation degree of $f$, up to some approximation error $\epsilon$. While determining the spectral measure is computationally non-trivial, assuming access to it allows us to derive tight information-theoretic lower bounds on query complexity. Once such optimal polynomials are found, it can be implemented via the Quantum Singular Value Transformation (QSVT) framework \cite{Gily_n_2019, Martyn_2021, Lin_2020}. 
Therefore, our work unifies quantum query models, orthogonal polynomial theory, and Krylov compression 
into a single coherent framework \cite{Koornwinder_2013, M_ck_2022, Balasubramanian_2022, Adhikari_2022}.

The remainder of this paper is organized as follows. 
In Section~\ref{sec:KrylovMethods}, we review the basics of Krylov compression, 
the Lanczos algorithm, and the Favard polynomial formalism. 
Section~\ref{sec:operational-picture} develops the operational picture of the overall framework. 
In Section~\ref{sec:optimal-query-construction}, we analyze the optimal quantum query complexity 
within the state-aware formalism, some examples, and comparisons to the state-oblivious approach. 
Section~\ref{sec:family} extends the framework to a family of initial states, and finally, Section~\ref{sec:discussions} summarizes the conclusions and outlines future work.

\section{Krylov Compression, Lanczos Recurrence, and Favard Polynomials}
\label{sec:KrylovMethods}

Let us consider a quantum system with a time-independent Hermitian Hamiltonian \(H\) on a Hilbert space \(\mathcal H\).
The time evolution of a state $\ket{\psi(t)}$ is given by a Schrödinger equation:
\begin{equation}
    i\partial_t \ket{\psi(t)} = H \ket{\psi(t)}
\end{equation}
with solution $|\psi(t)\rangle = e^{-iHt}|\psi_0 \rangle$. The solution has a power series expansion as:
\begin{equation}
    \ket{\psi(t)} = \sum_{n = 0}^\infty \frac{(-it)^n}{n!} \ket{\psi_n}, \quad \ket{\psi_n} =  H^n \ket{\psi_0}. 
\end{equation}
Thus, the entire time evolution is built from repeated applications of $H$ to the starting state. Applying the Gram--Schmidt procedure to the sequence $\{|\psi_n\rangle\} = H^n \ket{\psi_0}$, produces an ordered, orthonormal basis  
\begin{equation}
\mathcal{K} = \{ |K_n\rangle : n=0,1,2,\dots\},
\qquad |K_0\rangle = |\psi_0\rangle.
\end{equation}
The basis $\mathcal{K}$ is also referred to as the Krylov basis, and the dimension of the Krylov basis denoted by $m =  \dim \mathcal{K}(H,|\psi_0\rangle)$ is called the Krylov dimension.  Depending on the Hamiltonian and the choice of initial state, the Krylov dimension may be much smaller than the full Hilbert space dimension, since linear dependencies can arise among the vectors $H^n|\psi_0\rangle$. In this case, the dynamics of $|\psi(t)\rangle$ are effectively confined to this reduced subspace. If \(m\) is finite, all evolution remains in an \(m\)-dimensional space; otherwise \(m=\infty\).

\paragraph{Lanczos Algorithm}

Starting from $\ket{K_0}=\ket{\psi_0}$, such an orthonormal basis (and Krylov space) can be constructed recursively via the Lanczos algorithm: 
\begin{equation}
\ket{A_{n+1}} = \left(H - a_n\right)\ket{K_n} - b_n \ket{K_{n-1}}
\end{equation}
where
$ \ket{K_{n}} = b_{n}^{-1} \ket{A_{n}}$, and $b_0 = 0$. The $a_n$ and $b_n$ are called Lanczos coefficients
\begin{equation}
a_n = \bra{K_n} H \ket{K_n}, 
\qquad b_{n} = \sqrt{\langle A_n | A_n \rangle}.
\end{equation}
This recursion implies
\begin{equation}
\label{eq:Lanczos_recurrence}
H\ket{K_n} = a_n \ket{K_n} + b_{n+1}\ket{K_{n+1}} + b_n \ket{K_{n-1}}.
\end{equation}
Thus, in the Krylov basis, the Hamiltonian is a tridiagonal matrix with Lanczos coefficients $a_n$ on the diagonal and $b_n$ on the off-diagonal.
\begin{equation}
J = 
\begin{pmatrix}
a_0 & b_1 &        &        &   \\
b_1 & a_1 & b_2    &        &   \\
    & b_2 & a_2    & \ddots &   \\
    &     & \ddots & \ddots & b_{m-1} \\
    &     &        & b_{m-1} & a_{m-1}
\end{pmatrix}.
\end{equation}
The matrix $J$ is commonly referred to as a Jacobi or Hessenberg matrix; here we will refer to it as the Krylov matrix for convenience. 

This representation is equivalent to a one-dimensional tight-binding Hamiltonian or quantum walk model, with $a_n$ acting as on-site potentials and $b_n$ as nearest-neighbor hopping amplitudes. In this sense, the Lanczos algorithm maps the dynamics of a general Hamiltonian onto an effective one-dimensional chain, greatly simplifying its study. In the Krylov basis, the state evolves as
\begin{equation}
|\psi(t)\rangle \;=\; \sum_{n=0}^{m-1} \psi_n(t)\,|K_n\rangle,
\end{equation}
with amplitudes obeying
\begin{equation}
i\,\dot\psi_n(t) \;=\; b_n\,\psi_{n-1}(t) + a_n\,\psi_n(t) + b_{n+1}\,\psi_{n+1}(t),
\end{equation}
subject to \(\psi_{-1}(t)=\psi_m(t)=0\) and \(\psi_0(0)=1\). The mean position
\begin{equation}
C(t) \;=\; \sum_{n=0}^{m-1} n\,|\psi_n(t)|^2
\end{equation}
quantifies the spread of the wavefunction on the Krylov chain.  $C(t)$ is also called by Krylov complexity in the literature \cite{Balasubramanian_2022}, but we will refer to it as a mean position to avoid confusion with other notions of complexity.

\paragraph{Spectral measure}
The Lanczos coefficients can also be obtained recursively, via the method outlined in \cite{Balasubramanian_2022}, from the moments of the Hamiltonian in the initial state:
\begin{equation}
   M_k \;=\;      \langle \psi_0| H^k |\psi_0\rangle =  \langle K_0| H^k |K_0\rangle 
\end{equation}
The moment itself can also be obtained from the survival amplitude, which is the amplitude that the state at time $t$ is the same as at the initial state:
\begin{equation}
  S(t) =   \langle \psi(t) | \psi_0 \rangle   = \langle \psi(t) | K_0 \rangle  = \psi_0(t)^{*}  
\end{equation}
Survival amplitude is then just the moment-generating function for the Hamiltonian of the initial state:
\begin{align}
\left. \frac{d^n}{dt^n} S(t) \right|_{t=0}
&=\left. \langle \psi(0) | 
  \frac{d^n}{dt^n} e^{iHt} \ket{\psi(0)}
  \right |_{t=0}\nonumber \\
&= \bra{K_0} (iH)^n \ket{K_0} = M_k.
\end{align}

This is particularly appealing because the moment or survival amplitude method works even for the infinite-dimensional system, and can provide Lanczos coefficients and the corresponding Krylov basis.

In fact, several key dynamical quantities, including Lanczos coefficients, can directly be inferred from the spectral measure itself. 
The pair \((H,|\psi_0\rangle)\) admits a unique probability measure \(\mu\) supported on \(\mathrm{spec}(H)\) such that
\begin{equation}
\label{eq:spectral_measure}
\langle \psi_0| f(H) |\psi_0 \rangle \;=\; \int f(\lambda)\, d\mu(\lambda),
\end{equation}
for all bounded functions $f$. For the discrete/finite-dimensional spectra, the measure $d\mu(\lambda)$ is given by: 
\begin{equation}
    \frac{d\mu(\lambda)}{d \lambda} = \sum_{n} \delta(\lambda - \lambda_n)\, |\langle \lambda_n | \psi_0 \rangle|^{2},
\end{equation}
One can see the measure $\mu$ as a quantity that tells which eigenvalues of $H$ matter for the initial state $\ket{\psi_0}$, and by how much.
The moments are then given by
\begin{equation}
M_k \;=\; \langle \psi_0| H^k |\psi_0\rangle \;=\; \int \lambda^k\, d\mu(\lambda)
\end{equation}
while the survival amplitude turns out to be just the Fourier transform of $\mu$:
\begin{equation}
    S(t) = \langle \psi_0 | e^{-iHt} | \psi_0 \rangle 
= \int e^{-i\lambda t} \, d\mu(\lambda).
\end{equation}
with Taylor expansion that shows explicitly how the first few moments control the short-time dynamics
\begin{equation}
S(t) \;=\; \sum_{k=0}^\infty \frac{(-it)^k}{k!}\, M_k.
\end{equation}
The Laplace transformation of the measure $\mu$ gives the classical partition function:
\[
Z(\beta) = \langle \psi_0 | e^{-\beta H} | \psi_0 \rangle
= \int e^{-\beta \lambda} \, d\mu( \lambda).
\]
while the Stieltjes transformation gives the Green's function:
\begin{equation}
    G(z) = \langle \psi_0 | (z - H)^{-1} | \psi_0 \rangle
= \int \frac{1}{z - \lambda} \, d\mu( \lambda).
\end{equation}
The continued fraction of $G(z)$ also encodes the Lanczos coefficients $a_n$ and $b_n$ as:
\begin{equation}
G(z) = \cfrac{1}{z - a_{0} - 
  \cfrac{b_{1}^{2}}{z - a_{1} - 
    \cfrac{b_{2}^{2}}{z - a_{2} - \cdots}}}.
\end{equation}

\paragraph{Krylov polynomials}

The Lanczos recursion coefficients  $(a_n,b_n)$ define a three-term recurrence relation for a sequence 
of polynomials $\{P_n(\lambda)\}$:
\begin{equation}
\label{eq:Favard_Recurrence}
\lambda P_n(\lambda) 
= b_{n+1} P_{n+1}(\lambda) + a_n P_n(\lambda) + b_n P_{n-1}(\lambda).
\end{equation}
with $P_0(\lambda) = 1$, and $b_0 = 0$. By Favard's theorem, there exists a positive measure  $\mu$ on $\mathbb{R}$ such that the polynomials $\{P_n\}$ are  orthogonal with respect to $\mu$:
\begin{equation}
\int P_m(\lambda) P_n(\lambda)\, d\mu(\lambda) 
=  \delta_{mn}.
\end{equation}
This measure is precisely the spectral measure $\mu$ associated 
with the pair $(H,|\psi_0\rangle)$ introduced in Eq. \ref{eq:spectral_measure}. 
Via the spectral theorem, one can replace $\lambda$ with $H$ in Eq. \ref{eq:Favard_Recurrence}, and obtain: 
\begin{equation}
\label{eq:recurrence2}
H P_n(H) \;=\; b_{n+1} P_{n+1}(H) \;+\; a_n P_n(H) \;+\; b_n P_{n-1}(H).
\end{equation}
Acting with both sides on $\ket{K_0}$, and using $\ket{K_n} = P_n(H) \ket{K_0}$ (orthogonal to all lower Krylov vectors), one obtains: 
\begin{equation}
H \ket{K_n} \;=\; b_{n+1}\ket{K_{n+1}} \;+\; a_n \ket{K_n} \;+\; b_n \ket{K_{n-1}} 
\end{equation}
which is equivalent to the Krylov/Lanczos recurrence Eq. \ref{eq:Lanczos_recurrence}. The orthogonality of Krylov basis is expressed by:
\begin{equation}
\begin{aligned}
           \langle K_n \mid K_m \rangle
&= \langle \psi_0 \mid P_n(H) P_m(H) \mid \psi_0 \rangle \\
&= \int d\mu(\lambda)\, P_n(\lambda) P_m(\lambda) \\
&= \delta_{nm}.  
\end{aligned}
\end{equation}

Therefore, the Lanczos coefficients $(a_n,b_n)$, Krylov matrix $J$, the Krylov polynomials $\{P_n(\lambda)\}$ (From Favard's theorem)  and the spectral measure $\mu$, are several equivalent ways of encoding  the same structure. We refer to these orthogonal polynomials as Krylov polynomials. 

Furthermore, this orthogonal polynomial structure guarantees optimality of approximation in $L^2(\mu)$. 
Any square-integrable function $f$ admits an expansion
\begin{equation}
\label{eq:favardExpansion}
    f(\lambda) = \sum_{n=0}^\infty c_n P_n(\lambda),  
    \quad c_n = \int f(\lambda) P_n(\lambda)\,d\mu(\lambda).
\end{equation}
The coefficients $c_n$ can be computed via the Gaussian quadrature method, such that the integrals get replaced by summation. As this is a standard numerical procedure, we omit the details here.

The best degree-\(d\) approximation is obtained by projecting \(f\) onto the span 
of these polynomials. This projection has the form
\begin{equation}
\label{eq:FavardTruncation}
    p_d(\lambda) = \sum_{n=0}^d c_n P_n(\lambda)
\end{equation}
The residual $f-p_d$ is orthogonal to every polynomial of degree at most $d$, as all of its components along \(P_0, \dots, P_d\) vanish by construction. 

To see that $p_d$ is the unique best approximation in \(L^2(\mu)\), let $q$ be any other 
polynomial with degree at most $d$.

\begin{align}
\|f - q\|_{L^2(\mu)}^2 \notag
&= \int |f(x) - q(x)|^2 \, d\mu(x) \\ \notag
&= \int \big( (f(x) - p_d(x)) + (p_d(x) - q(x)) \big)^2 \, d\mu(x) \\ \notag
&= \|f - p_d\|_{L^2(\mu)}^2 
   + 2 \int (f(x) - p_d(x)) \\ \notag 
   & \text{      } (p_d(x) - q(x)) \, d\mu(x)
   + \|p_d - q\|_{L^2(\mu)}^2.
\end{align}
The cross term vanishes because \(f - p_d\) is orthogonal to every degree-\(d\) 
polynomial, including \(q - p_d\), and we obtain

\[
\|f - q\|_{L^2(\mu)}^2
  = \|f - p_d\|_{L^2(\mu)}^2 + \|p_d - q\|_{L^2(\mu)}^2.
\]
Since the second term is nonnegative, we obtain
\[
\|f - q\|_{L^2(\mu)}^2 \ge \|f - p_d\|_{L^2(\mu)}^2.
\]
Therefore \(p_d\) is the unique polynomial of degree at most \(d\) that minimizes 
the \(L^2(\mu)\) approximation error.

\section{Operational picture}
\label{sec:operational-picture}

In this section, we will show how Krylov compression from original full Hilbert space to effective Krylov space captures the necessary physics about the action of $H$ on $\ket{\psi_0}$. More precisely, the dynamics of $(H,\ket{\psi_0})$ are exactly reproduced by $(J,\ket{e_1})$. To see this, let us introduce an isometry
\begin{equation}
V:\mathbb C^m\to\mathcal H,\qquad V\ket{e_j}=\ket{k_j}\quad (1\le j\le m),
\end{equation}
where $\{\ket{e_j}\}$ is the standard basis of $\mathbb C^m$, and $m$ is the Krylov dimension for the pair $(H, \ket{\psi_0})$. Mathematically, $V^\dagger V=I_m$, $P:=VV^\dagger$ is the orthogonal projector onto $\mathcal K_m$ and the Krylov compression of $H$ to $\mathcal K_m$ is given by the Krylov matrix $J$ as
\begin{equation}
J \;:=\; V^\dagger H V \;\in\;\mathbb C^{m\times m}.
\end{equation} 

Operationally, in the state-aware formalism, $\mathbb{C}^m$ is the smallest Hilbert space containing all physically relevant information about the problem.  
If $m \ll \dim \mathcal{H}$, the number of qubits needed to reproduce the same physics is reduced from $\lceil \log_2 \dim \mathcal{H} \rceil$ to $\lceil \log_2 m \rceil$. A canonical example of this phenomenon is the glued trees problem of quantum walk where the dynamics is mapped from full Hilbert space to Krylov space of polynomial size. This exponential compression leads to quantum advantage in such quantum walk algorithms \cite{Childs_2003}.

\subsection{Intertwining relation}
A crucial mathematical reason why this compressions works is the following polynomial intertwining relation
\begin{equation}
    p(H)V = V p(J), \qquad \text{for all polynomials } p .
\end{equation}
This follows from $H^n V = V J^n$ for all $n \ge 0$. To see this, we can use the principle of induction. The case $n=0$ is trivial. For $n=1$,
\begin{align}
(HV)\ket{e_j} 
  &= H\ket{k_j} \nonumber \\
  &= \sum_{i=1}^m \ket{k_i}\braket{k_i|H|k_j} \nonumber \\
  &= \sum_i \ket{k_i} \,(V^\dagger H V)_{ij} \nonumber \\
  &= (VJ)\ket{e_j}.
\end{align}
Thus $HV = VJ$, and the general step follows from $H^{n+1}V = H(H^nV) = (HV)J^n = (VJ)J^n = VJ^{n+1}$.
By linearity, $p(H)V = Vp(J)$ for any polynomial $p$.
Furthermore, by polynomial approximation, this extends further to any function $f$
that can be uniformly approximated by polynomials on $\mathrm{spec}(H)$:
\begin{equation}
    f(H)V = V f(J).
    \label{eq:functional-intertwine}
\end{equation}
Applied to the initial state $\ket{\psi_0}$,  this yields
\begin{equation}
    f(H)|\psi_0\rangle = f(H)V|e_1\rangle = V f(J)|e_1\rangle .
\end{equation}
Thus, while $H$ and $J$ are not globally equivalent, their action on the Krylov subspace
generated by $|\psi_0\rangle$ is exactly the same. Consequently, any computation that
applies $f(H)$ to $|\psi_0\rangle$ can be carried out by applying $f(J)$ to $|e_1\rangle$
in the reduced $m$-dimensional space and mapping back with the isometry $V$.

\subsection{Ampltitudes and Correlators}
By applying the polynomial intertwining relation termwise to the power series
$e^{-iHt} = \sum_{n\ge 0}\frac{(-it)^n}{n!} H^n$, one obtains
\begin{equation}
    e^{-iHt}V = \sum_{n\ge 0}\frac{(-it)^n}{n!} H^n V
              = \sum_{n\ge 0}\frac{(-it)^n}{n!} V J^n
              = V e^{-iJt}.
    \label{eq:unitary-intertwine}
\end{equation}
In particular, if $\ket{\psi_0}=\|\psi_0\|\,V\ket{e_1}$, then the time-evolved state
obeys
\begin{equation}
    e^{-iHt}\ket{\psi_0} = V e^{-iJt}\ket{e_1}.
    \label{eq:state-equality}
\end{equation}
This ensures that Schr\"odinger evolution of states starting in $\mathcal{K}_m$ is exactly reproduced by evolution in compressed space $\mathbb C^m$.

However, for general observables $A$ that do not necessarily commute with the Krylov projector, the calculation becomes an approximation. For any observable $A$ with compression $\tilde{A} = V^\dagger A V$, the Heisenberg evolution satisfies:
\begin{align}
   \langle\psi_0|A(t)|\psi_0\rangle &=  \bra{e_1} V^\dagger e^{iHt} A e^{-iHt} V \ket{e_1} \\
   &\approx \bra{e_1} e^{iJt} (V^\dagger A V) e^{-iJt} \ket{e_1} \\
   &= \langle e_1|\,e^{iJt}\,\tilde{A}\,e^{-iJt}\,|e_1\rangle.
\end{align}
This approximation is exact only if $A$ maps the Krylov subspace to itself. More generally, for any $A_1,\dots,A_r$ and times $t_1,\dots,t_r$, we obtain the Krylov approximation:
\begin{equation}
\label{eq:heisenberg-multitime}
\braket{\psi_0|A_1(t_1)\cdots A_r(t_r)|\psi_0}
\;\approx\;
\bra{e_1} \prod_{s=1}^r \Big(e^{iJ t_s}\,\tilde A_s\,e^{-iJ t_s}\Big) \ket{e_1}.
\end{equation}
For example, two-time correlators are approximated by:
\begin{equation}
\braket{\psi_0|A(t)B(s)|\psi_0}
\;\approx\;
\bra{e_1} e^{iJ t}\tilde A\, e^{-iJ (t-s)} \tilde B\, e^{-iJ s} \ket{e_1}.
\end{equation} 
Therefore, the relevant amplitudes and correlators can be efficiently approximated in $\mathbb{C}^m$, provided the Krylov subspace captures the dominant support of the operators acting on the state.

Overall, this implies that one could instead compute the correlators using the compressed tri-diagonal Hamiltonian $J$, which is just a one-dimensional tight-binding Hamiltonian or a one-dimensional quantum walk problem. Starting with the standard basis $\ket{e_0}$, and computing the correlators of the operator $\tilde{A} = V^\dagger A V$ would give the same result as for the actual Hamiltonian $H$ and the state $\ket{\psi_0}$.

\section{ State-aware optimal query complexity: construction}
\label{sec:optimal-query-construction}

In this section, we study the query complexity of preparing, within some $\epsilon$-approximation, a state proportional to 
\begin{equation}
      \ket{\psi_{\mathrm{out}}} = f(H)\,\ket{\psi_0}
\end{equation}
where $H$ is a Hermitian operator (or its block-encoding) accessible only 
through queries to the corresponding unitary 
\begin{equation}
    U = e^{iH} \quad \text{or} \quad U_{\mathrm{block}} = 
    \begin{pmatrix} H & \cdot \\ \cdot & \cdot \end{pmatrix}.
\end{equation}
The query complexity, $q_\text{min}(\epsilon)$, is defined as the total number of such queries made by the algorithm. There exist other query models, such as the power query model used in Shor's period finding, but that falls outside the scope of our analysis.

A key observation is that any $q$-query algorithm can only generate states of
the form $p(H)\ket{\psi_0}$ where $p$ is a polynomial of degree at most $q$.
Thus, the cost of preparing $f(H)\ket{\psi_0}$ is entirely determined by the
minimal polynomial degree required to approximate $f$ on the spectral support
relevant to $\ket{\psi_0}$.  
We then define the state-aware degree functional as
\begin{equation}
\label{eq:nmu}
\begin{aligned}
n_\mu(f,\varepsilon)\;:=\;
\min\bigl\{\,n:\;&\exists\,p\ \text{with}\ \deg p\le n,\\[-2pt]
&\|f-p\|_{L^2(\mu)}\le \varepsilon \bigr\}.
\end{aligned}
\end{equation}
where,
\begin{equation}
\label{eq:state-error}
\begin{aligned}
\|f - p\|_{L^2(\mu)}^2 
&= \int |f(x)-p(x)|^2\, d\mu(x) \\
&= \bigl\|(f(H)-p(H))\ket{\psi_0}\bigr\|^2.
\end{aligned}
\end{equation} 
Operationally, small $\bigl\|(f(H)-p(H))\ket{\psi_0}\bigr\|$ implies no experiment can reliably distinguish the $f(H)\ket{\psi_0} $ state from the $p(H)\ket{\psi_0}$.

It gives us the information-theoretic lower bound for the minimum number of queries, $q_{\min}$, required as
\begin{equation}
\label{eq:lower-bound}
q_{\min}(\varepsilon)\;\ge\; n_\mu(f,\varepsilon).
\end{equation}
Among all such polynomials $p$, the error is minimized precisely by $p_d$, from equation \ref{eq:FavardTruncation}, as follows:
\begin{equation}
\begin{aligned}
\|f - p_d\|_{L^2(\mu)} \le \|f - p\|_{L^2(\mu)}, \\
\forall\, p \text{ with } \deg p \le d.
\end{aligned}
\end{equation}
One can implement $p_d$ with the QSVT framework (after rescaling), thus achieving the bound exactly with just $d$ queries. Hence
\begin{equation}
q_{\min}(\varepsilon)\;=\;n_\mu(f,\varepsilon).
\end{equation}
These results can be summarized in a single theorem as follows:
\begin{theorem}[Krylov--Favard Query Duality]
Let $H$ be a Hermitian operator and $|\psi_0\rangle$ an initial state with
spectral measure $\mu_{H,\psi_0}$. Construct the Krylov basis
$|K_n\rangle = P_n(H)|\psi_0\rangle$ via the Favard--Lanczos recurrence
\[
\lambda P_n(\lambda) = b_{n+1}P_{n+1}(\lambda) + a_n P_n(\lambda) + b_n P_{n-1}(\lambda),
\quad P_0(\lambda) = 1.
\]
For any bounded function $f$, its expansion in this intrinsic orthogonal
basis is
\[
f(H)|\psi_0\rangle = \sum_{n=0}^{\infty} c_n |K_n\rangle, \qquad
c_n = \int f(\lambda) P_n(\lambda)\, d\mu(\lambda).
\]
Let $p_d(H)|\psi_0\rangle = \sum_{n=0}^{d} c_n |K_n\rangle$ be the
degree-$d$ truncation. Then the minimal number of oracle queries required
to prepare $f(H)|\psi_0\rangle$ within error $\varepsilon$ equals the
smallest integer $d$ such that
\[
\| f(H)|\psi_0\rangle - p_d(H)|\psi_0\rangle \| \le \varepsilon.
\]
Equivalently,
\[
q_{\min}(\varepsilon) = n_{\mu}(f,\varepsilon),
\]
where $n_{\mu}(f,\varepsilon)$ is the minimal Favard degree satisfying the
above condition.
\end{theorem}

To validate this, we can follow the same argument as in Krylov polynomials. The action of $f(H)$ on $\ket{K_0}$ gives:
\begin{equation}
\begin{aligned}
    f(H)\ket{K_0} 
&= \sum_{n=0}^\infty c_n P_n(H)\ket{K_0} \\
&= \sum_{n=0}^\infty c_n \ket{K_n},
\end{aligned}
\end{equation}
where we used $\ket{K_n} \;=\; P_n(H)\ket{K_0}$. The action of $p_d(H)$ gives:
\begin{equation}
\begin{aligned}
    p_d(H)\ket{K_0} &= \sum_{n=0}^d c_n \ket{K_n} \\
    &= c_0 \ket{K_0} + c_1 \ket{K_1} + \dots + c_n \ket{K_n}.
\end{aligned}
\end{equation}
Since $ \bigl(f(H)-p_d(H)\bigr)\ket{K_0} 
= \sum_{n>d} c_n \ket{K_n}$, one obtains:
\begin{equation}
\begin{aligned}
  \bigl\|\,(f(H)-p_d(H))\ket{K_0}\,\bigr\|^2 &=   \|f-p_d\|_{L^2(\mu)}^2 \\
  &= \sum_{n>d} |c_n|^2. 
\end{aligned}
\end{equation}
Compared with arbitrary polynomial $q(\lambda)= \sum_{n = 0}^d a_n P_n (\lambda)$ with $\deg q \le d$,
\[
(f(H)-q(H))\ket{K_0}
= \sum_{n=0}^d (c_n-a_n)\ket{K_n} 
  + \sum_{n>d} c_n \ket{K_n},
\]
 the minimum is attained when $a_n=c_n$ for all $n\le d$, i.e.\ when $q=p_d$. The minimum error of degree-d truncation is given by $\sum_{n>d} |c_n|^2$:

\begin{equation}
\begin{aligned}
        n_\mu(f,\varepsilon)
&= \min\bigl\{\,d:\ \|f - p_d\|_{L^2(\mu)} \le \varepsilon\,\bigr\} \\
&= \min\bigl\{\,d:\ \|f - p_d\|^2_{L^2(\mu)} \le \varepsilon^2\,\bigr\} \\
& = \min\biggl\{\,d:\ \sum_{n>d} |c_n|^2 \le \varepsilon^2\,\biggr\}.
\end{aligned}
\end{equation}

\begin{algorithm}[H]
\label{algorithm1}
\caption{State-Aware Krylov Query Procedure}
\begin{algorithmic}[1]
\Require Hermitian $H$, initial state $|\psi_0\rangle$, target function $f$, error tolerance $\varepsilon$
\State Construct Krylov basis $\{|K_n\rangle\}$ via the Lanczos recurrence.
\State Compute Lanczos coefficients $(a_n,b_n)$ and the corresponding Jacobi matrix $J$.
\State Obtain spectral measure $\mu_{H,\psi_0}$ from $\{a_n,b_n\}$.
\State Expand $f$ in the Favard polynomial basis: $f(\lambda) = \sum_n c_n P_n(\lambda)$.
\State Truncate at degree $d = n_\mu(f,\varepsilon)$ such that $\|f - p_d\|_{L^2(\mu)} \le \varepsilon$.
\State Implement $p_d(H)$ using QSVT with $d$ oracle queries to $H$.
\Ensure Output state $\propto f(H)|\psi_0\rangle$ within error $\varepsilon$.
\end{algorithmic}
\end{algorithm}

 In particular, if the Krylov dimension $m$ is finite, i.e., $\mu$ has $m$ atoms,  then $n_\mu(f,0)$ equals the minimal interpolation degree of $f$ on those $m$ nodes. For a generic $f$, the query complexity is thus $m-1$. If the Krylov dimension is infinite, which can happen if the Hilbert dimension itself is infinite, then $\mu$ has infinite support, and one has to use approximations. In that case, query complexity equals $n_\mu(f,\varepsilon)$, where $\epsilon$ is the $L^2(\mu)$ approximation error. For example, for the time-evolution operator $f(\lambda)=e^{-i\lambda t}$, the Lanczos construction provides the optimal degree-$d$ polynomial  approximation to $e^{-iHt}|\psi_0\rangle$ within the Krylov subspace.

\subsection{Implementing using QSVT formalism}
In this section, we discuss how to implement the optimal polynomial $p_d(\lambda)=\sum_{n=0}^d c_n P_n(\lambda)$ using the QSVT formalism. For this, we have to express $p_d$ in Chebyshev form. Each Krylov polynomial can be written in the monomial basis as:
\begin{equation}
    P_n(\lambda)=\sum_{k=0}^n \alpha_{n,k}\lambda^k
\end{equation}
which gives the monomial expansion of $p_d$ as $p_d(\lambda)=\sum_{k=0}^d \beta_k\lambda^k$ with $\beta_k=\sum_{n=k}^d c_n\alpha_{n,k}$.

QSVT requires the operator argument to lie in the interval $[-1,1]$. Let $\alpha\ge\|H\|$ be a scaling factor. After rescaling the spectrum with $x=\lambda/\alpha$, we obtain the transformed polynomial
\begin{equation}
    q_d(x)=p_d(\alpha x)=\sum_{k=0}^d \hat\beta_k x^k
\end{equation}
 where $\hat\beta_k=\beta_k\alpha^k$. One can then write $q_d(x)$ in the Chebyshev basis 
 \begin{equation}
    q_d(x)=\sum_{j=0}^d \gamma_j T_j(x) 
 \end{equation}
 via standard basis-change techniques. 

 Let us also discuss the normalization issue. For $q_d(x)$ to be a valid QSVT polynomial, we require \(|q_d(x)| \le 1\) for all \(x \in [-1, 1]\). Let \(M = \max_{x \in [-1,1]} |q_d(x)|\) and the normalized polynomial \(\tilde{q}_d(x) = q_d(x)/M\). The coefficients of \(\tilde{q}_d\) are passed to a Quantum Signal Processing (QSP) phase-synthesis routine to generate the phase angles. The resulting circuit prepares the state proportional to \(f(H)\ket{\psi_0}\) with probability amplitude scaled by \(1/M\) using $O(d)$ QSVT queries. The success probability scales as $1/M^2$. To boost this probability close to unity, one can employ Amplitude Amplification (or Oblivious Amplitude Amplification for block-encodings), which introduces a multiplicative overhead of $\mathcal{O}(M)$ repetitions. Thus, the total query count is $\mathcal{O}(M \cdot n_\mu(f, \epsilon))$. For large $M$, it can sweep away any advantage gained through Krylov polynomial method. 

To control this overhead, we have two options. First, we can replace the standard Favard projection with a constrained optimization strategy. For a fixed maximum allowable scaling factor $M_{\max} \ge 1$, we solve:
\begin{equation}
    \label{eq:constrained_optimization}
    \min_{p \in \mathbb{P}_d} \|f - p\|_{L^2(\mu)} \quad \text{subject to} \quad \max_{x \in [-1, 1]} |p(x)| \le M_{\max}.
\end{equation}
This formulation ensures that the Amplitude Amplification cost is strictly bounded by $M_{\max}$, balancing the trade-off between the query depth $d$ and the success probability.

A second option, which is better suited for Krylov subspace, is to construct a Christoffel filter $\mathcal{F}_{\text{filter}}(x)$, and obtain a regularized approximation $\tilde{p}_{d'}(x)$ as 
\begin{equation}
    \tilde{p}_{d'}(x) = p_d(x) \cdot \mathcal{F}_{\text{filter}}(x).
\end{equation}
such that
\begin{equation}
   |\tilde{p}_{d'}(x)| \ll 1 \quad \forall x \notin \mathrm{supp}(\mu).
\end{equation}
thereby reducing the global maximum $M$ significantly. Although this increases the total query complexity to $d+ \deg(\mathcal{F})$, it drastically reduces the amplitude amplification overhead, often yielding a superior total complexity.

\subsection{Example: State-aware HHL}
\label{subsec:state-aware-hhl}

We now illustrate the state-aware formalism in the context of the quantum
linear systems problem (HHL algorithm). Take $f(x)=1/x$ and
$\mu=\mu_{A,b}$, the spectral measure of $(A,b)$ with occupied support
$S_b=\mathrm{supp}(\mu)$. The algorithm needs to act only on eigenmodes
populated by $\ket{b}$. The optimal query count equals the minimal $L^2(\mu)$ degree
required to approximate $1/x$ on $S_b$,
\begin{equation}
q_{\mathrm{opt}} = n_\mu(1/x,\varepsilon).
\end{equation}
If the occupied support lies in a continuous interval
$S_b\subseteq[\lambda_{\min}^{\mathrm{eff}},\lambda_{\max}^{\mathrm{eff}}]$,
then approximation theory yields
\begin{equation}
q_{\mathrm{opt}}=\Theta\!\big(\kappa_{\mathrm{eff}}\log(1/\varepsilon)\big),
\quad
\kappa_{\mathrm{eff}}=\lambda_{\max}^{\mathrm{eff}}/\lambda_{\min}^{\mathrm{eff}}.
\end{equation}
These effective endpoints can be substantially better than the global
condition number if $\ket{b}$ has little overlap on small eigenvalues.

\subsection{Generalization to a Family of Initial States}
\label{sec:family}

So far, we have focused on a single fixed initial state $\ket{\psi_0}$. In many applications, however, one wishes to process not just one input but a family of possible states
\[
\mathcal{S}=\{\ket{\psi^{(1)}},\ket{\psi^{(2)}},\dots,\ket{\psi^{(r)}}\}
\]
using the same Hamiltonian $H$. The relevant subspace in this setting is the joint Krylov space
\begin{equation}
\mathcal{K}_{\mathrm{fam}}(H,\mathcal{S}) \;=\;
\mathrm{span}\Bigl(\bigcup_{j=1}^r \mathcal{K}(H,\ket{\psi^{(j)}})\Bigr),
\end{equation}
where each $\mathcal{K}(H,\ket{\psi^{(j)}})$ is the cyclic Krylov subspace generated from $\ket{\psi^{(j)}}$. Let $m_{\mathrm{fam}}:=\dim\mathcal{K}_{\mathrm{fam}}(H,\mathcal{S})$. Clearly $m_{\mathrm{fam}}\leq\sum_{j=1}^r m_j$ where $m_j=\dim\mathcal{K}(H,\ket{\psi^{(j)}})$, and in practice $m_{\mathrm{fam}}$ can be much smaller whenever the individual subspaces overlap. Compressing $H$ to this joint space yields the $m_{\mathrm{fam}}\times m_{\mathrm{fam}}$ Jacobi surrogate $J_{\mathrm{fam}}$ that exactly reproduces the dynamics of $(H,\mathcal{S})$.

In this setting the optimal query complexity is determined by the family-aware degree functional, defined analogously to \eqref{eq:nmu} but with respect to the joint spectral measure supported on $\mathcal{K}_{\mathrm{fam}}$. For measures with finite support (i.e., when $\mathcal{K}_{\mathrm{fam}}$ closes after $m_{\mathrm{fam}}$ steps), one has
\begin{equation}
Q_{\mathrm{opt}}^{\mathrm{fam}} \;=\; m_{\mathrm{fam}}-1,
\end{equation}
matching the single-state case. For infinite support, the minimal query count is instead given by
\begin{equation}
Q_{\mathrm{opt}}^{\mathrm{fam}} \;=\; n_{\mu_{\mathrm{fam}}}(f,\varepsilon),
\end{equation}
the smallest degree-$n$ polynomial that approximates $f$ to accuracy $\varepsilon$ in $L^2(\mu_{\mathrm{fam}})$, where $\mu_{\mathrm{fam}}$ is the family spectral measure.

This shows the distinction from the worst-case setting. In the input-agnostic case, query complexity is governed by the degree of the minimax (Chebyshev) polynomial approximating $f$ on $\mathrm{spec}(H)$, which typically scales as $\mathcal{O}(1/\varepsilon)$ regardless of the input state. By contrast, the state-aware family formalism replaces this worst-case degree with the joint Krylov dimension $m_{\mathrm{fam}}-1$ (or more generally the family-aware degree functional), which depends only on the occupied spectral support of the family. When $\mathcal{S}$ overlaps only a restricted or highly correlated part of the spectrum, this can be orders of magnitude smaller than the worst-case scenario.

\section{Discussions and Conclusion}
\label{sec:discussions}

We have shown that in the state-aware setting, quantum simulation reduces exactly to evolution in the Krylov subspace generated by $(H,\ket{\psi_0})$, with the Jacobi matrix $J$ providing the minimal effective Hamiltonian. For the finite support case, and a generic function $f$, the Krylov dimension $K_D$ sets the optimal query complexity, while for the infinite support case, the optimal polynomial approximation follows from the Krylov/Favard approximations. This framework applies to any $f(H)$, including algorithms like HHL, and yields potentially large savings over worst-case methods. Extensions to Christoffel filtering, non-Hermitian settings, and quantum–classical advantage boundaries offer promising directions for future work, which are as follows.

A major practical limitation of our proposal is the assumption that the spectral measure $\mu_{H,\psi_0}$ is known in advance. Without this information, our result is  mostly information-theoretic rather than directly implementable. Determining $\mu$ efficiently therefore remains an significant open problem. In principle, the spectral measure can be estimated on a quantum device. Quantum Phase Estimation (QPE) samples eigenvalues according to $\mu$, while moment-estimation or survival-amplitude techniques access its Fourier transform and enable reconstruction of the associated Lanczos coefficients. In the future, we plan to develop scalable algorithms for learning $\mu_{H,\psi_0}$, and convert our framework into a practical tool as well.

Another limitation of our framework is that the final state could have reduced probability amplitude due to the normalization issues in QSVT. This could be solved via constrained opitmization strategy or using Christoffel filtering with the later being more suited for orthogonal polynomials theory. In fact, other techniques from the theory of orthogonal polynomials, such as spectral transformations (e.g., Christoffel filtering), classical inequalities, and results like Bochner’s theorem, can be leveraged to further reduce query complexity.

Another promising direction is on the boundary of quantum and classical computational advantage. So far, we have focused on the quantum query complexity. However, if the quantum query complexity is polynomial, while the classical methods would require superpolynomial or even exponential queries, the quantum-classical boundary can be defined. A similar approach has already been used in the context of quantum versus classical walks \cite{Childs_2003, balasubramanian2024exponentialspeedupsquantumwalks, adhikari2023quantuminformationspreadingscrambling}.

Krylov methods could also be used for tackling coherent noise errors. In Krylov-based algorithms, the spectral measure $\mu_{H,\psi_0}$, determined entirely by the initial state $|\psi_0\rangle$, fixes the orthonormal polynomial sequence $\{p_n\}$ and the recurrence coefficients $(\alpha_n, \beta_n)$ that govern the Krylov dynamics. Consequently, even a small coherent perturbation to $|\psi_0\rangle$ can significantly alter this measure, modifying the optimal polynomial approximation problem faced by the algorithm. Coherent implementation errors $H \mapsto H + \delta H$ induce inhomogeneous perturbations $(\alpha_n, \beta_n) \mapsto (\alpha_n + \delta \alpha_n, \beta_n + \delta \beta_n)$; sufficiently irregular variations act as disorder on the Krylov chain and can trigger Anderson localization, whereby amplitudes fail to reach large $n$ and quantities such as $C(t)$ saturate prematurely. This leads to a distinct, “silent” failure mode in which the circuit continues to operate, but computation stalls in low-$n$ sectors. Mitigation of such errors seems extremely crucial as well for a functioning quantum computation.

Finally, we note a significant implication of the Krylov-based state-aware formalism for Quantum Machine Learning (QML). A major bottleneck in variational quantum algorithms is the Barren Plateau phenomenon, where gradients vanish exponentially due to the exploration of the exponentially large Hilbert space~\cite{McClean_2018}. By restricting the problem to the Krylov subspace \(\mathcal{K}_m\), the effective dimension of the state manifold is reduced from \(2^N\) to \(m\). This could provide insights towards solving the Barren Plateau issues.

\section{Acknowledgments}

This research is part of the Munich Quantum Valley, which is supported by the Bavarian state government with funds from the
Hightech Agenda Bayern Plus.

\bibliography{apssamp}

\appendix
\end{document}